\documentclass[letterpaper, 10 pt, conference]{ieeeconf}

\IEEEoverridecommandlockouts         

\usepackage{amssymb}
\usepackage{graphicx}
\usepackage{hyperref}
\usepackage[utf8]{inputenc}

\usepackage[usenames,dvipsnames]{xcolor}
\usepackage{array}
\usepackage{flushend}

\usepackage{listings}
\newcommand{\lstinputSeparatedFlag}[5] {
    \vspace{2.5ex}
    \begin{ownfigure}
    \setlength{\parskip}{1ex plus0pt minus0pt}\noindent%
    \smash{\rlap{\hspace{0.637ex}\raisebox{-0.0ex}{\rightline{\scriptsize
    \tt \frame{\makebox(60, 11){#2}}}}}}%
    \lstset{language=#1}
    \lstinputlisting[xleftmargin=10pt, label=lst:#3, caption={#4}]{#5}
    \end{ownfigure}
}

\newcommand{\lstinput}[4] {
    \lstinputSeparatedFlag{#1}{#1}{#2}{#3}{#4}
}

\newenvironment{ownfigure}[0]%
{\begin{figure}[htb!]}
{\end{figure}}

\usepackage{color}
\definecolor{DarkRed}{rgb}{0.75,0,0}
\definecolor{Lightgreen}{rgb}{0.588,1.0,0.588}
\definecolor{DarkGreen}{rgb}{0,0.5,0}
    
\lstdefinelanguage{MontiArc}[]{Java}{
  morekeywords={component, port, in, out, inv, package, import, connect, autoconnect}
}

\lstdefinelanguage{myJava}[]{Java}{
  commentstyle=\color{DarkGreen}\itshape 
}

\lstdefinelanguage{MontiArcAutomaton}[]{Java}{
  morekeywords={component, port, in, out, inv, package, import, connect,
  autoconnect, automaton, state, ocl, java, initial, final, noCompletion,
  chaosCompletion, var}, commentstyle=\color{DarkGreen}\itshape }

\lstdefinelanguage{MCConfig} { 
    morekeywords={config, Require, Model} 
}

\lstdefinelanguage{Manifest} { 
    morekeywords={Manifest, Bundle, ManifestVersion, Name, SymbolicName,
      Version, Require
    } 
}

\lstdefinelanguage{mcGrammar}[]{}{
  morekeywords={
    grammar, package, path, parser, lexer, nows, noslcomments, nomlcomments, 
    noident, nostring, noanything, nocharvocabulary, dotident, identrule,
    xmlcomments, hashcomments, texcomments, freemarkercomments, concept, 
    globalnaming, define, usage, options, true, false, protected, ident, 
    compilationunit
  }
}

\lstdefinelanguage{mcLng}[]{}{
  morekeywords={
    dsltool, language, package, path, parser, root, parsingworkflow, 
    rootfactory, lexer, nows, noslcomments, nomlcomments, noident, nostring,
    dotident, concept, globalnaming, define, usage, options, true, false, 
    protected, ident
  }
}

\lstdefinelanguage{mcManifest}[]{}{
  morekeywords={
    bundle, Bundle, Name, SymbolicName, true, false, Main, Class, 
    Version, Activator, Localization, Require, 
    Exclude, Eclipse, LazyStart, Vendor, Export, Package, 
    ClassPath
  }
}

\lstdefinelanguage{Alloy}[]{Java}{
commentstyle=\color{DarkGreen}\itshape,
  morekeywords={abstract,sig,->,fact,pred,fun,run,for,iff,
  not,no,one,all,some,lone,\#,set,in,and,or,but,exactly,none,univ,Int,assert,check},
  otherkeywords = {[2]????},
    morekeywords = {[2]????},
    keywordstyle={[2]\color{blue}},
    otherkeywords = {[3]????,<,<->,->, &, |, =, !=, !,<:,~},
    morekeywords = {[3]????,<,<->,->, &, |, =, !=, !,<:,~},
    keywordstyle={[3]\color{blue}}
  }

\lstdefinelanguage{mccd}[]{Java}{
  morekeywords={classdiagram,abstract,<<singleton>>,class,int,String,
  association,composition,extends}
}

\lstdefinelanguage{FreeMarker}[]{}{
  keywordsprefix={\#},
  keywords={in},
  commentstyle=\color{DarkGreen}\itshape }

\lstdefinelanguage{Mona}[]{}{
  morekeywords={ex0,all0,ex1,all1,ex2,all2,var0,var1,var2,pred,in,notin,include,union,inter,empty,assert},
  morecomment=[l]{\#},
  commentstyle=\color{DarkGreen}\itshape,
  otherkeywords = {[2]????,next,boolean,init,case,esac},
  morekeywords = {[2]????,next,boolean,init,case,esac},
  otherkeywords = {[3]????,<,<=>,=>, &, |, =, !=, !},
  morekeywords = {[3]????,<,<=>,=>, &, |, =, !=, !},
}

\lstdefinelanguage{myPython}[]{Python}{
  morekeywords={assert},
  morecomment=[l]{\#},
  commentstyle=\color{DarkGreen}\itshape,
}

\lstdefinelanguage{Action}[]{Java}{
  morekeywords={action,parameters,execution,entry,exit},
  commentstyle=\color{DarkGreen}\itshape,
}

\usepackage{xspace}

\newcommand{\code}[1]{\texttt{#1}}
\newcommand{\concept}[1]{\textit{#1}}

\newcommand{\lr}{LightRocks\xspace}

\newcommand{\umlp}{{\mbox{UML/P}}\xspace}

\title{\LARGE \bf Modeling Reusable, Platform-Independent Robot
Assembly Processes}

\author{
    \authorblockN{Arvid Butting$^1$, Bernhard Rumpe$^1$, Christoph
    Schulze$^1$, Ulrike Thomas$^2$, and Andreas Wortmann$^1$} 
    \authorblockA{
    $^1$ Software Engineering, RWTH Aachen University, Germany\\
    $^2$ Robotics and Human-Machine Interaction, Technical University Chemnitz, Germany }
\thanks{This research has partly received funding from the European Union Seventh Framework Programme (FP7/2007-2013) under grant agreement n¡ 287787, SMErobotics.}
}

\lstdefinelanguage{UML/P Statechart}{
  morekeywords={statechart, initial, state, final}, 
  commentstyle=\color{DarkGreen}\itshape 
}

\begin{document}

\maketitle
\thispagestyle{empty}
\pagestyle{empty}

\begin{abstract}

Smart factories that allow flexible production of highly individualized
goods require flexible robots, usable in efficient assembly lines.
Compliant robots can work safely in shared environments with domain
experts, who have to program such robots easily for arbitrary tasks. We
propose a new domain-specific language and toolchain for robot assembly
tasks for compliant manipulators. With the \lr toolchain, assembly tasks
are modeled on different levels of abstraction, allowing a separation of
concerns between domain experts and robotics experts: externally
provided, platform-independent assembly plans are instantiated by the
domain experts using models of processes and tasks. Tasks are comprised
of skills, which combine platform-specific action models provided by
robotics experts. Thereby it supports a flexible production and re-use
of modeling artifacts for various assembly processes.

\end{abstract}

\section{Introduction}
\label{sec:Introduction}
Future smart factories require flexible production of highly 
individualized goods in small and medium lot sizes, where 
reconfigurations of systems occur with high frequencies. To achieve 
this, such factories require flexibly usable assembly line robots, which 
can work safely in shared environments with humans. Flexible usage of 
such robots requires easy programming for arbitrary tasks. Compliant 
manipulators allow such interaction, but programming these robots is 
more complex than programming rigid industrial robots. Beside knowledge 
of a general-purpose programming language (GPL), it also requires 
control specific knowledge to adjust compliance parameters like 
stiffness and damping for each motion. Thus, only robotics experts with 
sufficient programming knowledge are able to program such compliant 
manipulators. Furthermore, the re-usability of such control specific 
programs is endangered due to their target specific nature. Making these 
usable and re-usable in daily flexible production requires tools for a 
more abstract development of assembly tasks that are executable by shop 
floor workers with none software engineering expertise. 

In \cite{THR+13} we propose a new framework which consists of a 
domain-specific language (DSL) and a toolchain to generate specialized 
robot programs for assembly tasks. The recent robot programming 
interface for the iiwa-LBR is based on the compliance-frame concept 
introduced by Mason and further developed with the task-frame-formalism 
by~\cite{deSchutter88}. The programming interface uses stiffness and 
damping definitions for each Cartesian DOF while the robot moves towards 
a target pose. Stop conditions can be applied in order to immediately 
stopping the motion and awaiting new specifications for the next motion. 
Fig.~\ref{fig:SkillIntro} illustrates the programming specification 
necessary for the action shown at the right side of the figure. The 
robot rotates about the x-axis in the task-frame, while it pushes the 
object towards the top hat rail until the torque about the x-axis 
exceeds a certain threshold. For Cartesian force controlled robots the 
skill-primitive concept has been suggested earlier~\cite{HST92,TBW03}, 
which is also based on the task-frame formalism. Skill primitives (SPs) 
are combined to skill primitive nets~\cite{TFKW03,KFTW04,MMTW09} (SPNs) 
and their transitions are ensured by preconditions and postconditions. 
Figure~\ref{fig:SkillPrimitiveNet} shows a SPN consisting of three SPs, 
which describes the placement of an object onto a table. The surface 
orientation of the plane is not precisely known. Hence, a force-torque 
sensor is attached to the robot hand flange. Therewith contact forces 
and torques can be measured and evaluated during the assembly process. 
According to contact forces and torques, different SPs are selected for 
execution. In this example, either a rotation along the object's depths 
axis or its width axis is carried out. When the stop conditions trigger, 
a transition to one of the subsequent SPs, depending on the measured 
values of the force-torque sensor, is performed. 

\begin{ownfigure}
\centering
  \includegraphics[width=0.48\textwidth]{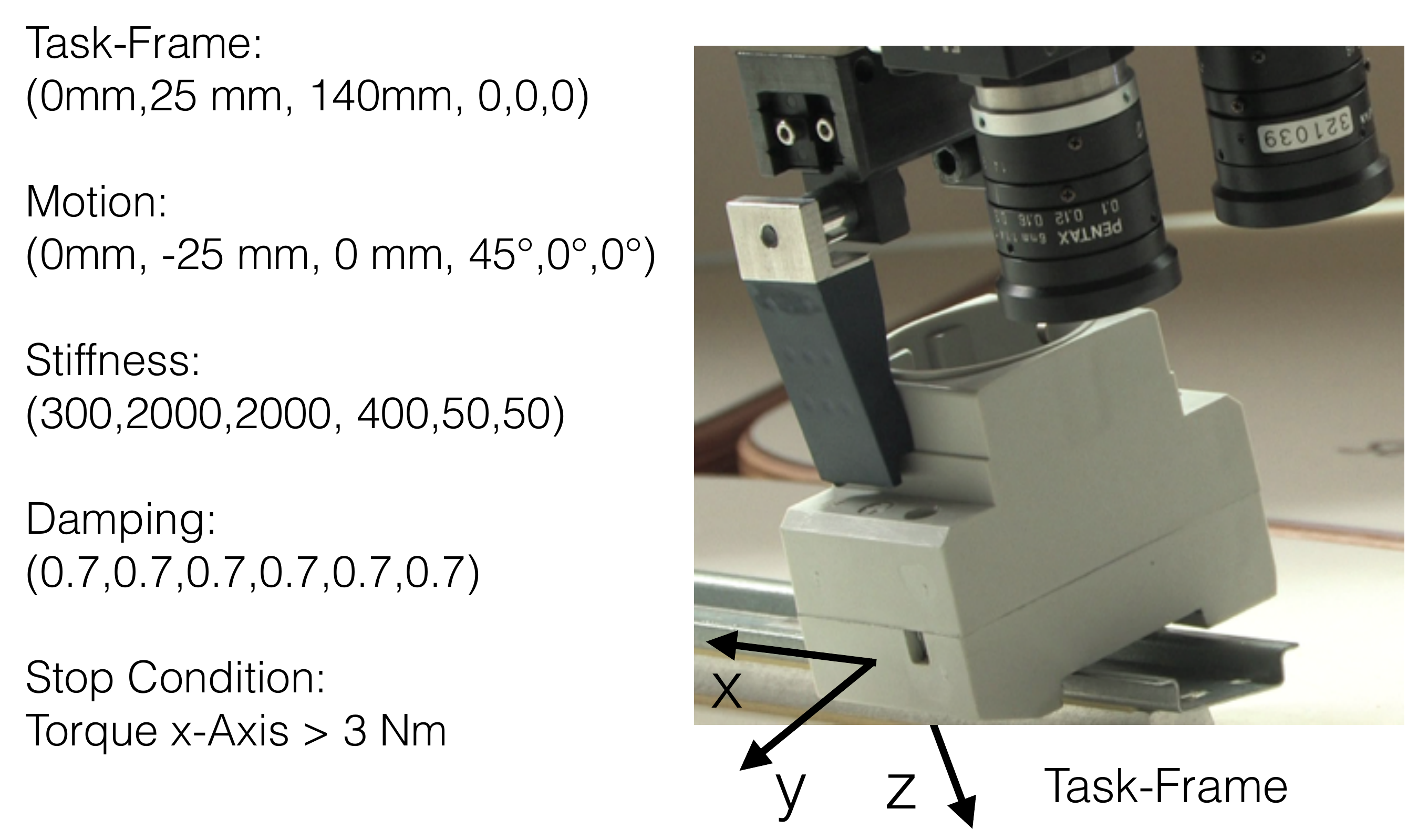}
  \caption{Left side: Motion commands for the LBR-robot to establish the 
  task. Right side: Position of the task frame for the rotation about the 
  x-axis to assemble the socket onto the top hat rail.}
  \label{fig:SkillIntro}
\end{ownfigure}

\begin{ownfigure}
\centering
  \includegraphics[width=0.48\textwidth]{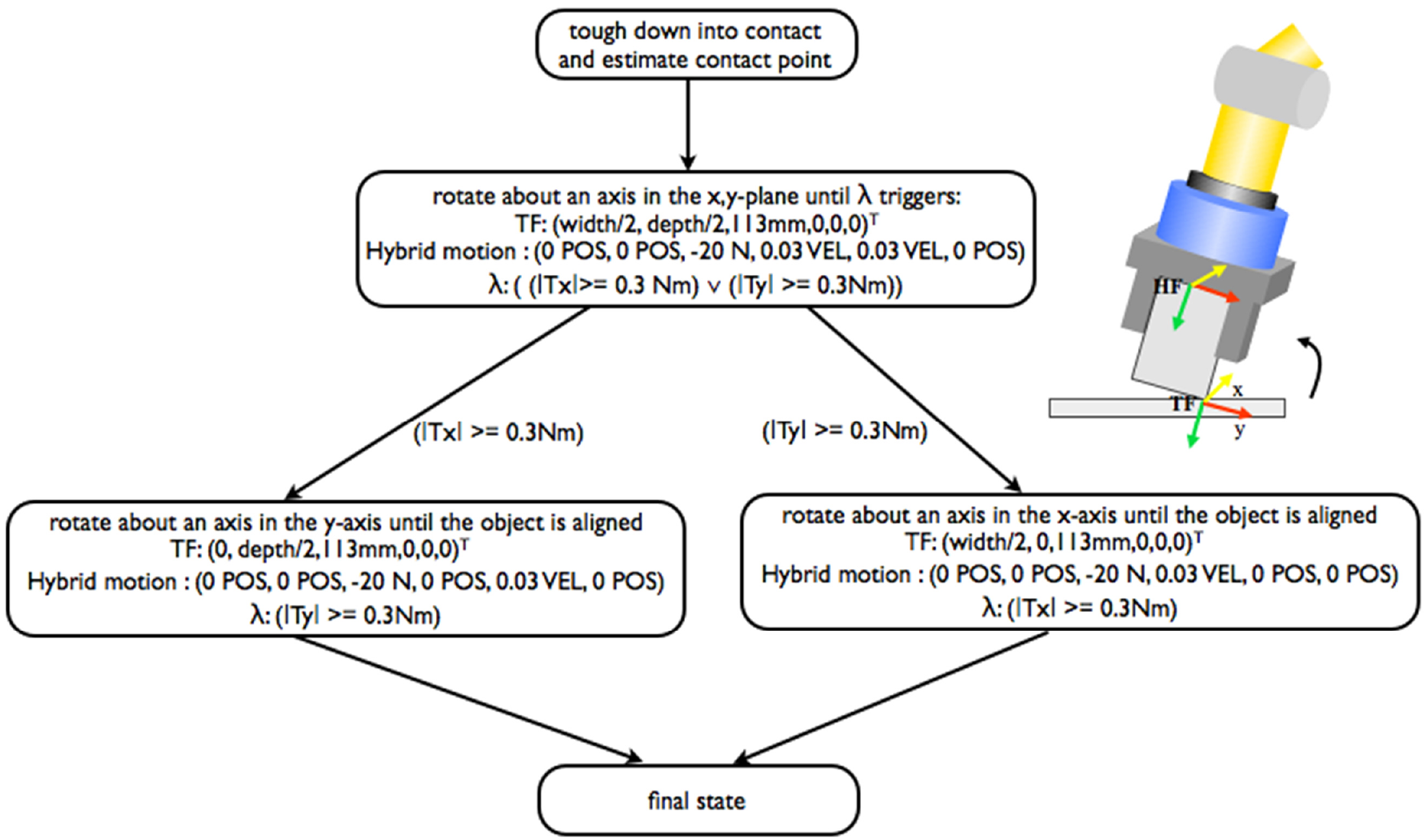}
\caption{Excerpt of a SPN for placing an object onto a table
despite uncertainties about the orientation of table's plane.}
\label{fig:SkillPrimitiveNet}
\end{ownfigure}

SPNs have proven useful for the programming of complex robot tasks but 
lack abstraction, separation of concerns between shop floor workers and 
robotics experts. As it can be seen, the art of programming different 
robots in the assembly domain, might be the same. Hence a concept is 
necessary which is grounded on domain-specific languages and allows the 
programming of different robots in an intuitive way triggered by the 
problem domain rather than the robot hardware. It motivated us to 
provide such a framework which allows experts to exploit robotic 
hardware while non-experts can use their results in an intuitive way.

Figure~\ref{fig:layers} illustrates how robot assembly tasks are modeled 
as three-level networks with \lr~\cite{THR+13} instead. We have adapted 
the concept of SPs such that it can be used for compliant manipulators. 
In order to increase the level of abstraction we distinguish between 
assembly \concept{plans}, \concept{processes}, \concept{tasks}, 
\concept{skills}, and \concept{actions}: Assembly plans are provided 
externally~\cite{Thomas08} and consist of assembly processes. Each 
assembly process consists of tasks that consist of skills. Assembly 
processes and tasks can be modified by domain experts to adjust the 
assembly process to fit new environmental conditions, e.g., because the 
robot uses another gripper than assumed, a sensor does not work as 
expected, or workpieces are placed differently than the expert system 
assumed. Assembly skills consist of actions, which are platform specific 
and provided by robot experts. 

\begin{ownfigure}
\centering
  \includegraphics[width=0.48\textwidth]{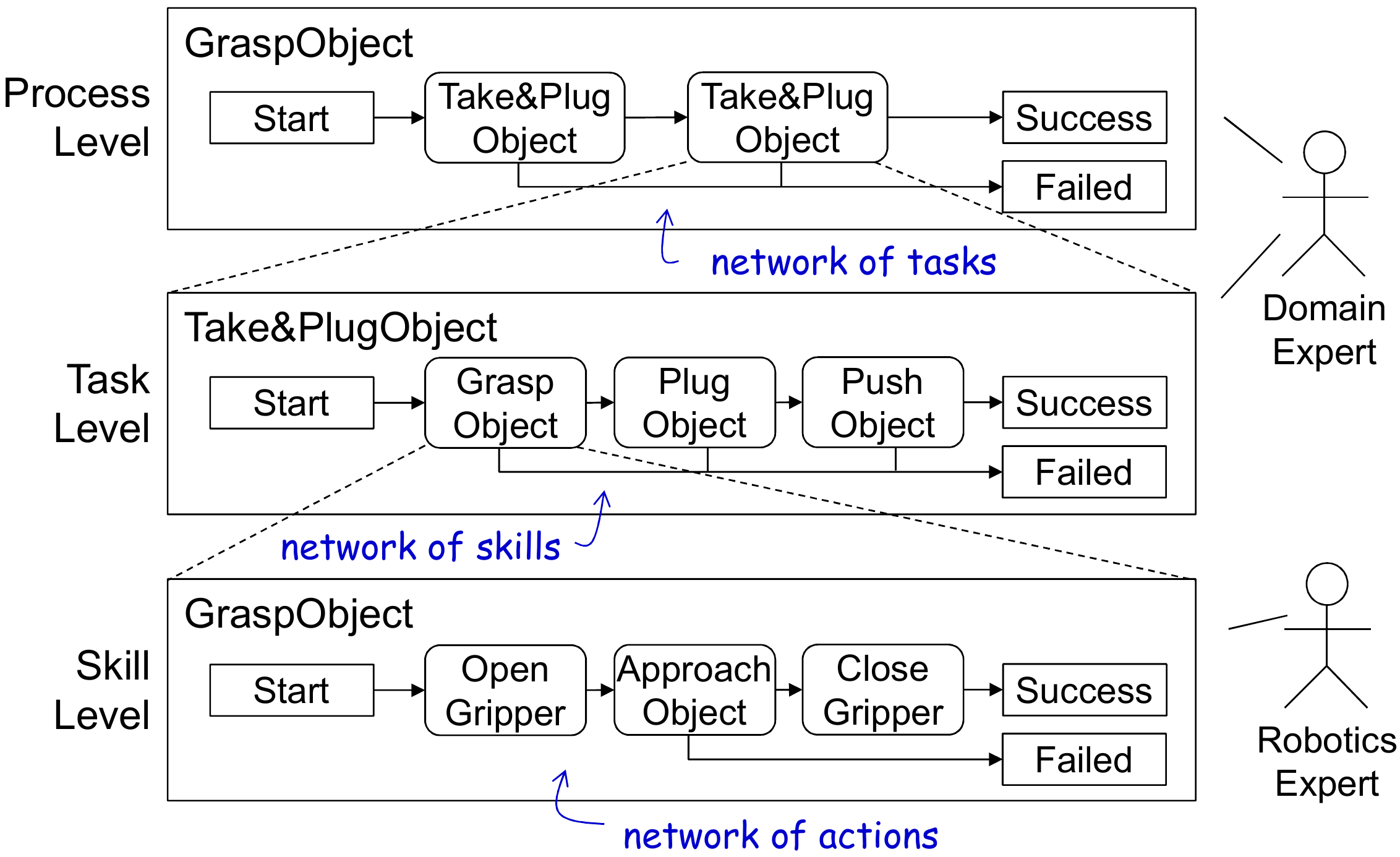}
\caption{The \lr abstraction layers.}
\label{fig:layers}
\end{ownfigure}

These levels of abstraction allow to separate robotics expertise modeled 
within actions and skills from domain expertise embodied in tasks and 
processes. It also allows re-using recurring skills for different 
assembly tasks and recurring tasks for different assembly processes. The 
previous version of \lr~\cite{THR+13} was implemented as a profile of 
the \umlp~\cite{Rum11} Statechart (SC) language with the MontiCore 
language workbench~\cite{GKR+06,KRV10}. The \umlp is a variant of 
UML~\cite{OMG10} for programming. While this allows re-use of language 
infrastructure, description of domain types via \umlp class diagram (CD) 
models~\cite{LNPR+13}, model analyses, and code generators~\cite{Sch12}, 
the resulting modeling language is less comprehensible than intended and 
requires domain experts to comprehend the full expressiveness of \umlp SCs. 
To liberate domain experts from this, we present a collection of MontiCore 
DSLs for concise representation of SPNs to facilitate development of 
re-usable, platform-independent robot assembly tasks. In addition, the 
degree of re-usability for each introduced abstraction layer regarding 
different tasks, APIs, target platforms or robots is evaluated. 

In the following, Sect.~\ref{sec:Example} illustrates \lr by example 
before Sect.~\ref{sec:RelatedWork} discusses related work. 
Sect.~\ref{sec:LR2} describes the new \lr DSLs and toolchain. 
Afterwards, Sect.~\ref{sec:CaseStudies} outlines case studies and 
finally, Sect.~\ref{sec:Conclusion} debates future work and summarizes 
the contribution. 

\section{Example}
\label{sec:Example}

The assembly task of placing a screw into a thread may be composed by
grasping the screw, moving it to the thread, and tightening it into the
thread. Figure~\ref{fig:GraspAndScrew} depicts a part of the task
\code{GraspAndScrew} that uses the gripper to pick up a screw and
tightens it into a thread. The task consists of several skills and
provides two outcomes: either the screw is placed accordingly to the
skill \code{Screwing} or it is not. The skill \code{Screwing} inserts a
screw into a thread after a previously executed skill placed it
accordingly. It consists of four actions of which \code{Spin} and
\code{CloseGripper} are illustrated in Fig.~\ref{fig:GraspAndScrew}.

\begin{ownfigure}
\centering
  \includegraphics[width=0.48\textwidth]{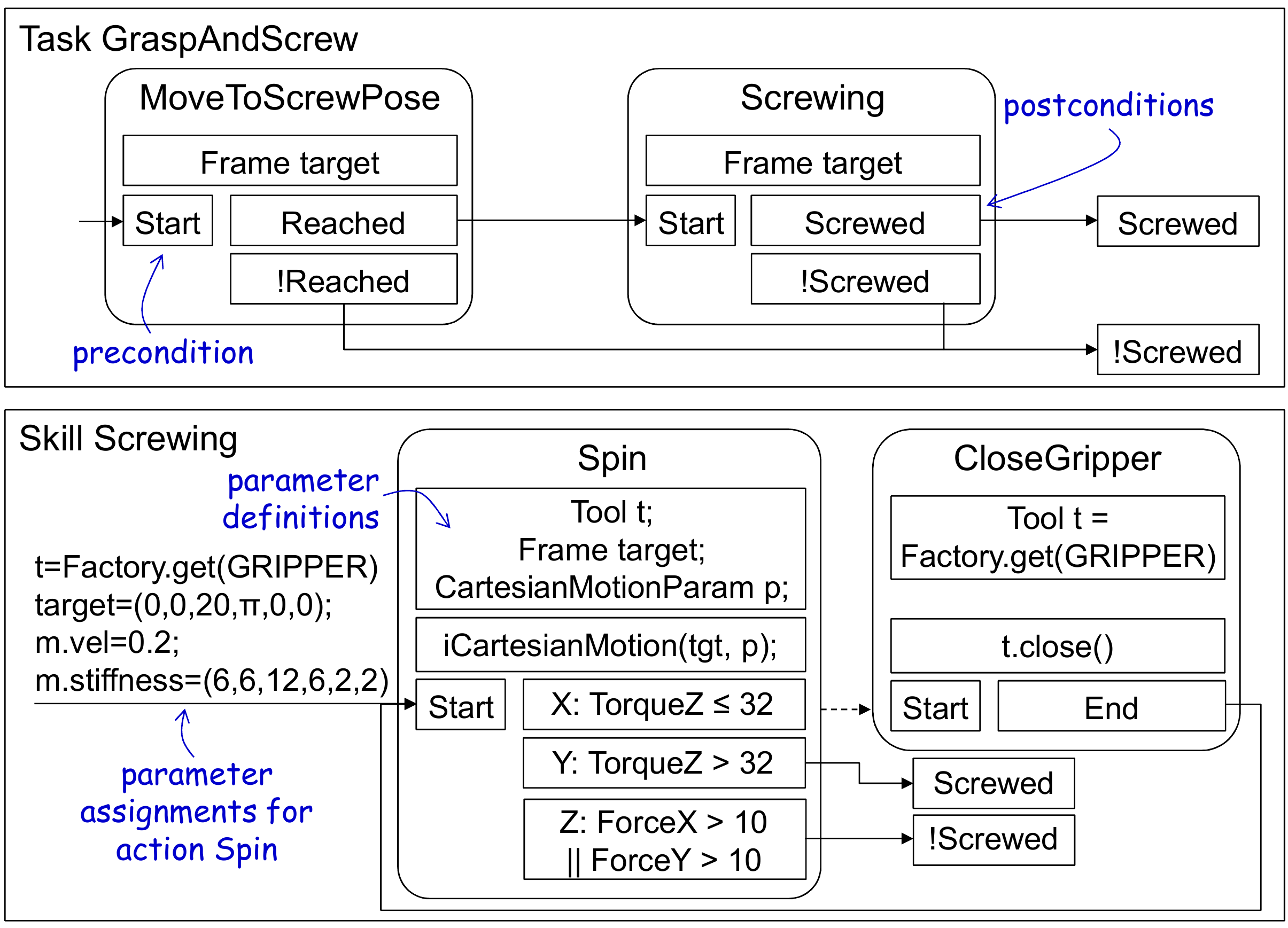}
\caption{Excerpts of task \code{GraspAndScrew} and skill
\code{Screwing}.}
\label{fig:GraspAndScrew}
\end{ownfigure}

The developer modeled the skill following human behavior: after
initially grasping the screw and holding it over the thread, the robot
spins the screw (action \code{Spin}). If a certain torque is reached,
the screw is fixed and the skill finishes. Otherwise the robot releases
the gripper, rotates back, grasps the screw again (cf. action
\code{CloseGripper}), and spins it again. Note that the action
\code{Spin} yields two further outcomes to detect whether something
besides the robot manipulated the workpiece. Such multiple outcomes
allow modeling flexible skills that can deal with uncertainties.
The action \code{CloseGripper} references the type \code{Tool}, which is
part of the robot interface of the domain model, via \texttt{t.close()}.

\lr uses MontiCore to validate such models and to generate proper GPL
implementations of these models. The resulting robot behavior is
illustrated in Fig.~\ref{fig:ResultingBehavior}.

\begin{ownfigure}
\centering
  \includegraphics[width=0.48\textwidth]{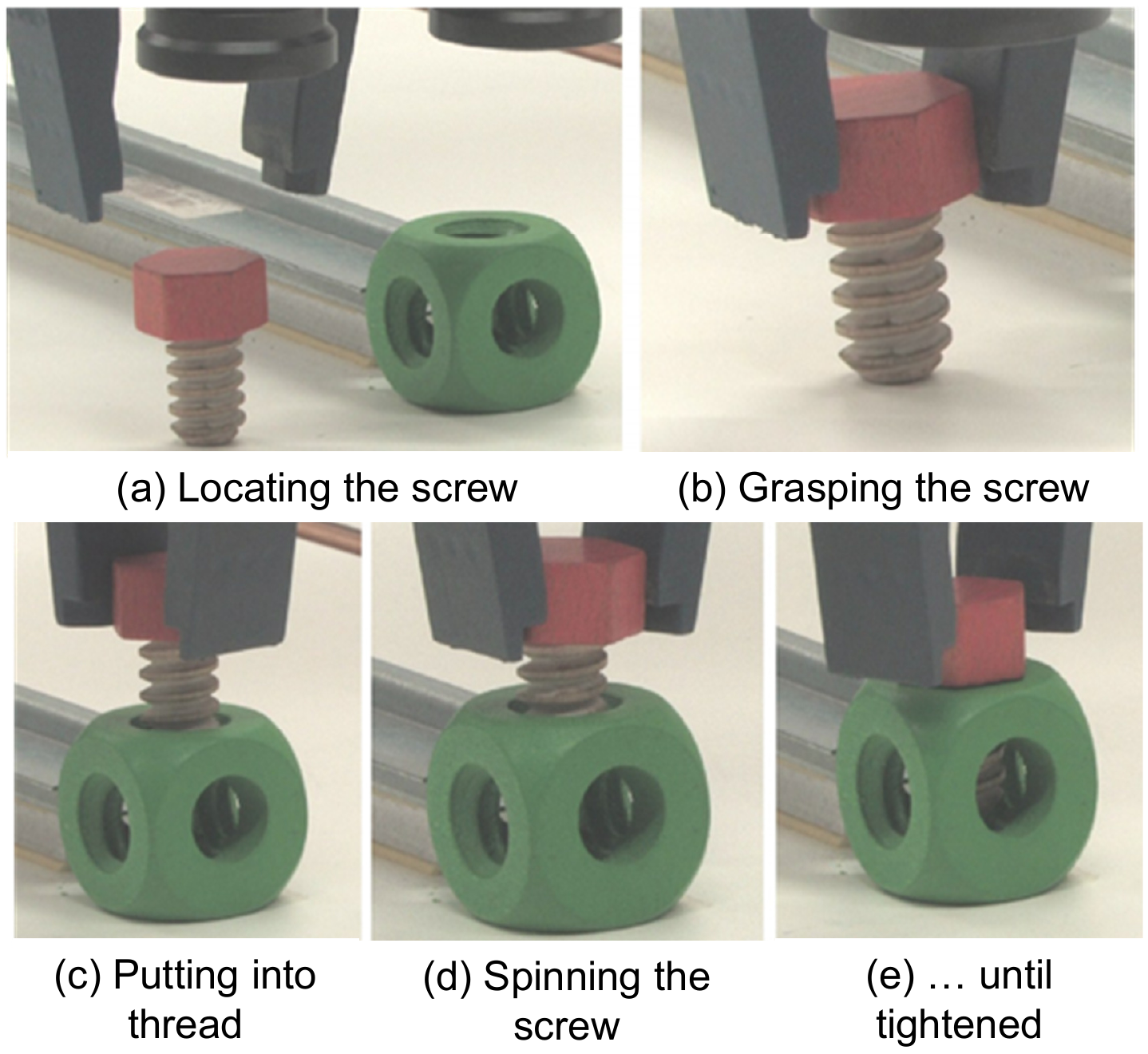}
\caption{A KUKA LBR robot finding and tightening a screw using
\lr}
\label{fig:ResultingBehavior}
\end{ownfigure} 

\section{Related Work}
\label{sec:RelatedWork}  

Recently, multiple DSLs for imperative or event-driven robot
behavior~\cite{KAH+10,BDH+10,ASH+12}, perception tasks~\cite{HSV+13},
and software architecture with state-based behavior
descriptions~\cite{SSL11,KSB+11,RRW13ATPS} have been presented.
While not focused on assembly tasks, these behavior modeling languages
are related in the common aim to facilitate robot programming.

Current approaches to robot behavior modeling aim at robotics software
engineers, not domain experts. To this end, these approaches either
provide less abstraction \cite{KAH+10,BDH+10,ASH+12} or require
knowledge on automata semantics to describe
behavior~\cite{SSL11,KSB+11,RRW13ATPS}. Even previous \lr~\cite{THR+13}
and closely related approaches~\cite{BBH+13,VKS+13} expect certain
degrees of software engineering knowledge from the domain experts. With
current \lr, this is relieved further as well-formedness rules prohibit
tasks and skills to reference the robot's API. Thus, domain experts only
need to comprehend task composition, skill composition and domain
parameter assignment.

Rethink developed an industrial robot called Baxter which can be trained
manually by ordinary line workers~\cite{Fit13}. Different kind of tasks, like
performing a blind pick or placing an object in a grid pattern, can
be configured by interacting with the UI and teaching positions and areas by
moving the robot's end effector directly. As described in~\autoref{subsec:KUKAAssembly} LightRocks supports
manual teaching of different poses, too.  Unlike LightRocks the approach
provides a user-friendly UI to define tasks only, while the proposed skill level
is hidden to the end user and consequently new skills can not be defined by the
customer directly.

\section{LightRocks Languages and Toolchain}
\label{sec:LR2}

\lr is developed as an integrated collection of MontiCore~\cite{KRV10} 
languages which comprises of a process language, a task language, a 
skill language, and an action language. In this collection, each 
language comprises of (i) a context-free grammar (CFG)~\cite{KRV07}, 
(ii) well-formedness checks, to check properties not expressible with 
CFGs, and (iii) symbol tables~\cite{Voe11,LNPR+13}, which give 
information about imported models of the same and other languages to 
enable language integration. The latter, for instance, enable 
integration with \umlp CDs to reference the models representing domain 
types and type checking between models. Models of \lr are 
textual~\cite{GKR+07}, which allows easy comprehension even of large 
models by software engineering experts, liberates these from layouting 
efforts, and enables processing models with their accustomed tools. 

Processes and tasks represent the logical structure of re-usable assembly
process knowledge. To this effect, processes contain a net of tasks and
tasks contain a net of skills. Both may refer only to domain types.
Skills contain nets of actions. Actions, however, may reference a single
method of the underlying domain model only. The domain model describes
the interfaces of sensors, tools, the types of movement that can be used
(e.g., linear or rotational), and the available domain types in terms of
a \umlp CD language profile that restricts class diagrams to interfaces.
Using different CDs for domain types and robot interfaces separates
concerns and enables to re-use both with arbitrary manipulators, as long
as a code generator from CD to the manipulator's GPL is present.
Ultimately, it grounds the assembly processes via actions to the robot
and allows validating the models.

We implemented previous \lr as a profile of \umlp SC where
tasks, skills, and actions were handled uniformly as states. This led to
``notational noise''~\cite{Wil01}, e.g., unintuitive language elements,
and increased the ``accidental complexity''~\cite{FR07} by forcing
domain experts to learn SCs instead of using an assembly domain
specific language. The new stand-alone \lr languages clearly separate
between language elements for domain experts and language elements for
robotics experts. This reduces noise and complexities.
Listing~\ref{lst:Spin} shows the textual model of the action \code{Spin}
as depicted in Fig.~\ref{fig:GraspAndScrew}.

\lstinput
{Action}
{Spin}
{Model of the action \code{Spin} as show in Fig.~\ref{fig:GraspAndScrew}.} 
{listings/Spin.action}

The textual syntax is straightforward and consists of the keyword
\code{action} followed by a name (l.~1), a \code{parameters} declaration
block (ll.~3-7), which defines how the action can be parametrized, an
\code{execution} block (ll.~9-11) that may reference the robot API, and
a set of entry and exit rules (ll.~13-16) to define preconditions and
postconditions of the action. Parameters are assigned via incoming
transitions and locally visible in task, skill, or action.
Tasks and skills additionally propagate their parameters to the
contained topology.

\begin{figure}
\centering
  \includegraphics[width=0.48\textwidth]{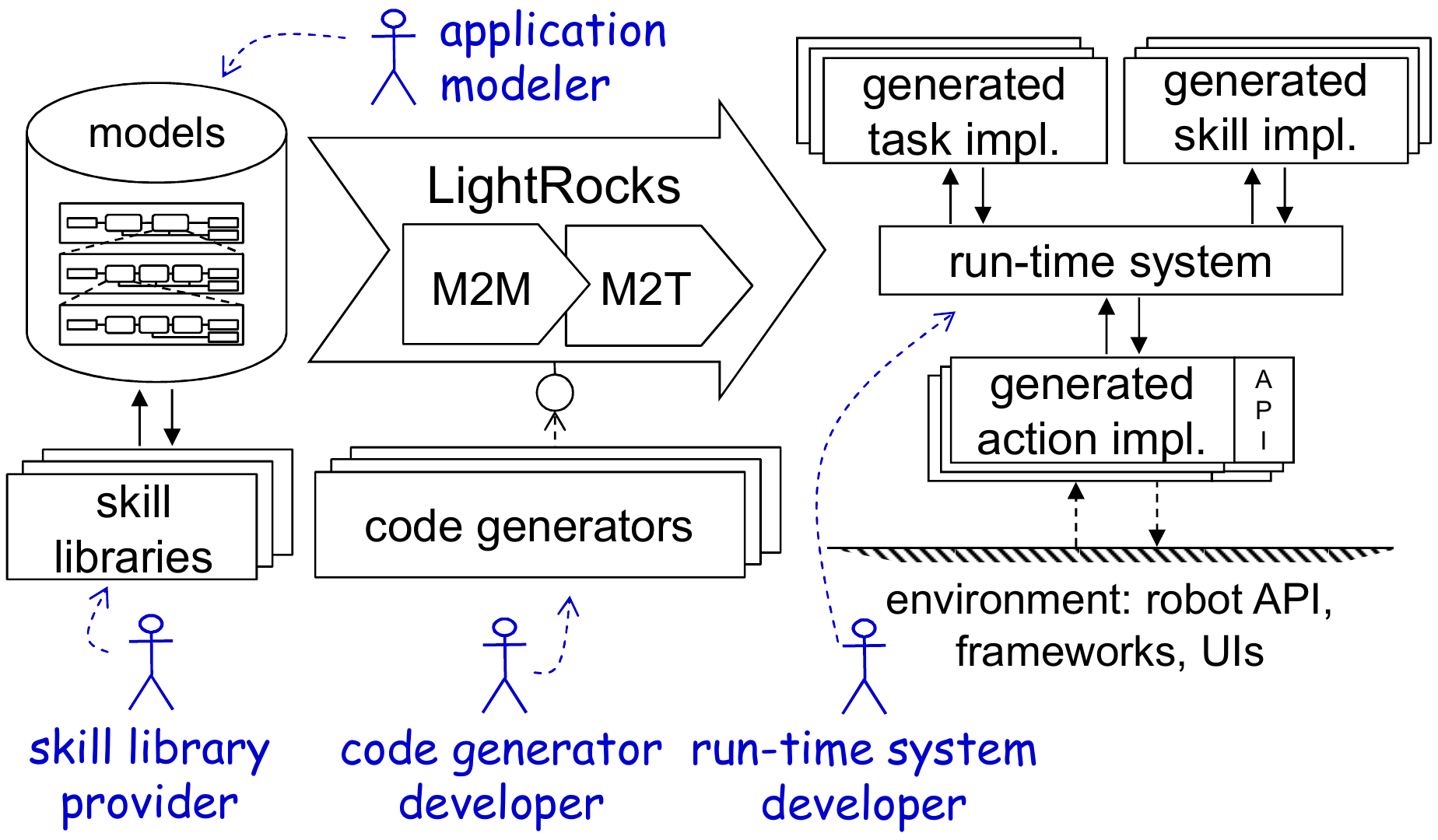}
\caption{The \lr toolchain translates assembly processes consisting
of tasks, skills, and actions via Statecharts to executable
code.}
\label{fig:Toolchain}
\end{figure}

Based on the grammars of \lr languages, MontiCore generates language
processing infrastructure including FreeMarker-based\footnote{FreeMarker
template engine: \url{http://www.freemarker.org}} code generation
sub-frameworks and text editors~\cite{KRV07a}. The \lr toolchain
utilizes these to parse models, process these, and ultimately transform
these into executable code.
Figure~\ref{fig:Toolchain} depicts the toolchain with the related roles,
according to~\cite{KRV06}.

As current and previous \lr are functionally equivalent, retaining
compatibility with existing models, tooling, and code generators is
straightforward: task, skill, and action models are transformed into
SC representations compatible with previous \lr which can be
processed by existing tooling. The current \lr toolchain
parses process models provided by the \emph{application modeler}.
Robotics experts acting as \emph{skill library providers} may provide
the latter. Current code generators retain the structural separation into
tasks, skills, and actions, which interact with a run-time system that,
for instance, defines how to execute transitions. The \lr toolchain
provides extension points for code generators to enable code generation
for arbitrary target platforms with minimal effort. Code generators and
run-time system (RTS) are not specific to a certain robot but to the
programming language to be used and non-functional requirements (e.g.,
restrictions to the amount of memory to be used). Thus, \emph{code
generator developers} and \emph{RTS developers} require software
engineering expertise, but no robot expertise. Platform-independence of
code generators and RTS ultimately enables their re-use and further
facilitates robot task development with \lr.

We also have developed a combined graphical and textual editor for
convenient modeling of assembly tasks by factory floor workers, domain
experts, and robotics experts. The editor provides two views: one for
modeling and one for model execution. It further allows parallel textual
and graphical modeling, parsing, and well-formedness checking of tasks
and their constituents. The editor is built with MontiCore's text editor
generation features, hence the text editor itself is generated: a
corresponding editor grammar allows to define keywords, outline
elements, filters, and other features from which text editor plugins for
Eclipse\footnote{Eclipse project:
\url{http://www.eclipse.org/}} are generated. The graphical editor is
also implemented as an Eclipse plugin and uses the Standard Widget
Toolkit\footnote{SWT website: \url{http://www.eclipse.org/swt/}} (SWT)
to render tasks, skills, and actions.

Figure~\ref{fig:ExecutorView} shows the editor's execution view with the
textual editor top middle and the graphical editor bottom middle.
The graphical editor displays the currently edited network of tasks
parallel to the corresponding textual model. Model parsing, context
condition checks, outline and syntax highlighting of the text editor are
directly re-used from the \lr toolchain. The model execution framework
takes care of monitoring and representing the currently executed parts
of the model. The contents of textual and the graphical editor are
synchronized directly, by either informing the textual or the graphical
editor about any modifications of the model. Once the developer starts
modeling, the graphical editor invokes MontiCore to either parse the
changed textual model or prints the changed model into the displayed
text editor. Layout data are stored separately and do not pollute the
textual model.

\begin{ownfigure}
\centering
  \includegraphics[width=0.48\textwidth]{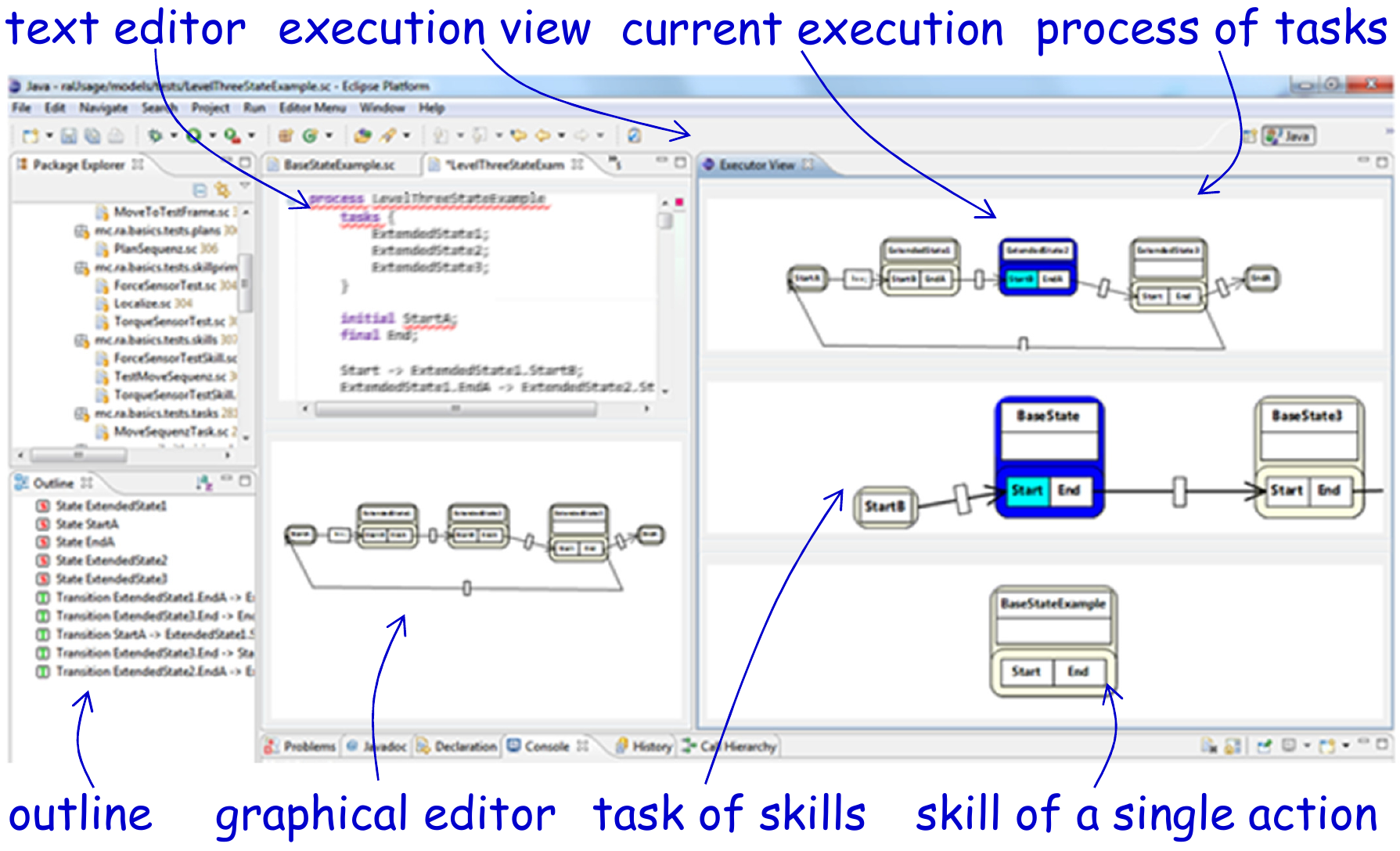}
\caption{The model execution view shows the currently executed action
and its parents parallel to graphical and textual editors.}
\label{fig:ExecutorView}
\end{ownfigure}

The right part of the editor shows the three assembly process levels and
their execution states at run-time. The editor highlights the currently
executed action and its parents. The top section shows the process level
and highlights the active task, the middle section shows the task level
and highlights the active skill, and the bottom section shows the skill
level and highlights the active action. Currently, the editor does not
support on-line editing of \lr models at run-time. While desirable, we
yet need to ascertain the requirements on valid run-time changes and the
implications on error handling mechanisms.

\section{Case Studies}
\label{sec:CaseStudies}

We have evaluated \lr with KUKA LBR robots and Lego Mindstorms robots.
With the former, we modeled classical assembly tasks. Due to hardware
restrictions of the Lego robots, we examined whether modeling of
non-assembly tasks is feasible with \lr. To our satisfaction, modeling
of re-usable logistics tasks with these robots was straightforward as
well. The following sections briefly report on both case studies.

\subsection{KUKA LBR Assembly Tasks}
\label{subsec:KUKAAssembly}  

First case studies were performed as typical assembly tasks with a KUKA
LBR manipulator. These modeled tasks included screwing, picking,
stacking, plugging, and different kinds of movements (e.g., force
controlled). The domain model for this case study comprised of $13$
interfaces for various concepts of domain and robot.
Figure~\ref{fig:UseCaseStackBlocks} shows the LBR successfully stacking
blocks as modeled with \lr. The process modeled to grasp and stack
a tower of four blocks consists of one task, which in turn consists of
four skills of between one and six actions.

\begin{ownfigure}
    \centering
    \includegraphics[width=.40\textwidth]{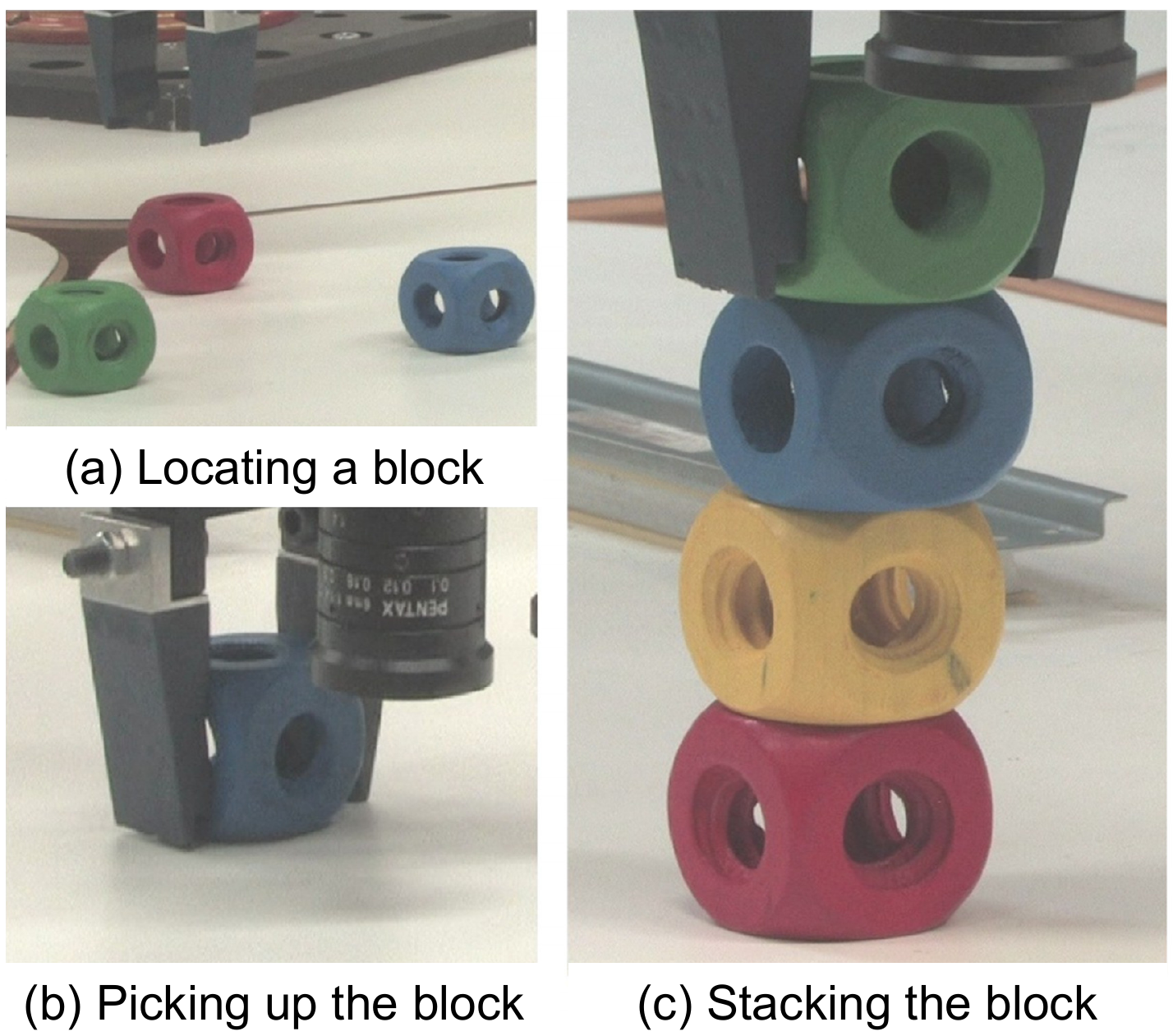}
    \caption{A LBR robot stacking colored blocks.}
    \label{fig:UseCaseStackBlocks}
\end{ownfigure}

Another typical assembly process is plugging workpieces onto another. We
therefore modeled a task for the LBR to plug safety sockets on a top-hat
rail~\cite{THR+13}. Figure~\ref{fig:UseCaseCapRail} shows the
performance of the LBR. The executed process model consists of a single
task that is repeated once per safety socket. The task itself consists
of five skills of up to five actions. Modeling assembly processes for
the LBR was straightforward and we could re-use tasks, skills, and
actions intuitively. We also observed that most re-use took place on
skill level, where - from a human perspective - simple behavior was
composed from actions. Skills regarding movement and grasping were
re-used most often.

\begin{ownfigure}
    \centering
    \includegraphics[width=.48\textwidth]{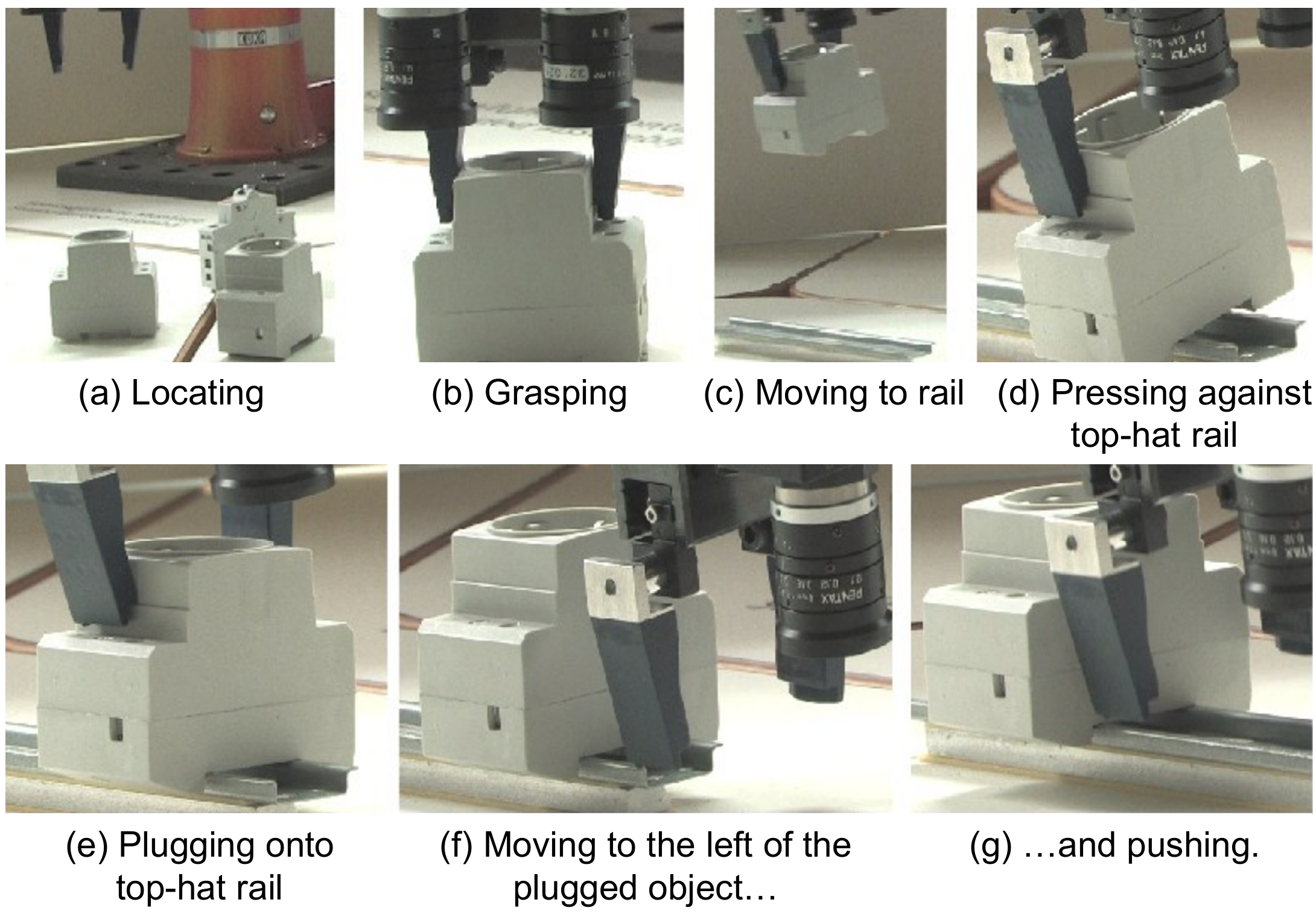}
    \caption{A LBR plugging safety sockets on a top-hat rail.}
    \label{fig:UseCaseCapRail}
\end{ownfigure}

\subsection{Lego NXT Logistics Tasks}

We also deployed \lr to Lego Mindstorms NXT robots to evaluate its usage
in different use cases. The Lego robots are designed for education and
easy access to robotics. Consequently, their hardware is restricted: out
of the box, there are neither laser scanners, nor compliant
manipulators. As \lr is not tied to platforms providing such hardware,
we designed a clean up scenario and modeled the processes accordingly.
In this scenario, a robot explores a fixed area while searching for
colored blocks. Whenever a block is detected, it is gripped and
collected in a container. The robot consists of a base with four wheels
and a manipulator (Fig.~\ref{fig:NXT}). The base uses a light sensor to
ensure moving within the defined area's boundaries, a front-mounted
distance sensor to detect blocks, and a manipulator to collect blocks.
The manipulator uses a color sensor to detect the blocks' colors.

\begin{ownfigure}
    \centering
    \includegraphics[width=.48\textwidth]{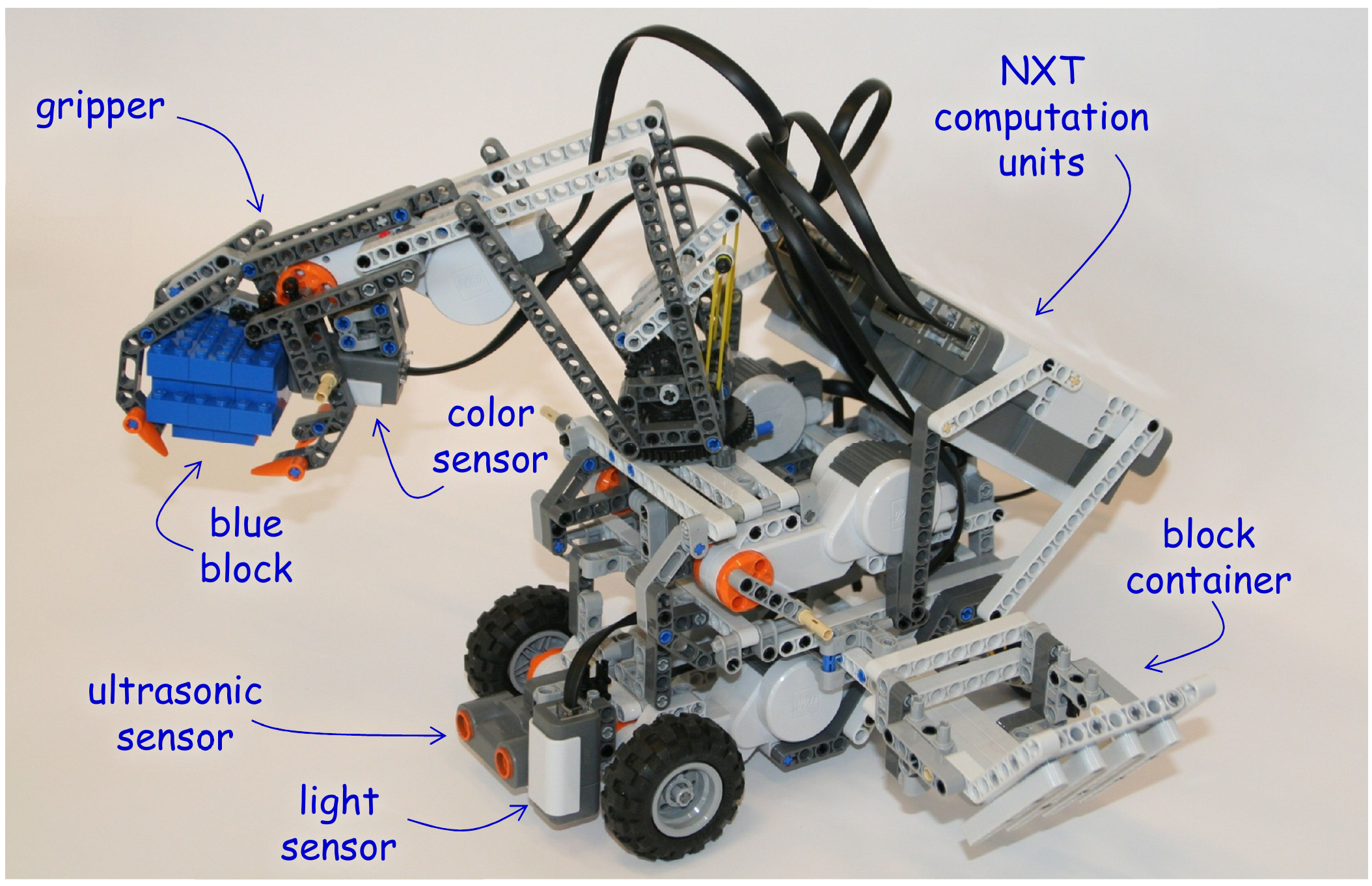}
    \caption{The Lego Mindstorms robot to perform clean up tasks.}
    \label{fig:NXT}
\end{ownfigure}


The \lr process to clean up colored blocks consists of three tasks, six
skills, and $14$ actions. Figure \ref{fig:ProcessCollectBlueObjects}
depicts the structure of the process \code{CollectBlueObjects}, which
contains a task \code{LookForObjects}, and a skill
\code{DriveToNextObject}. Skills and actions interface robot hardware
via the leJOS Java operating system\footnote{leJOS NXJ website:
\url{http://www.lejos.org/}} as robot API.

Due to lack of memory on the Mindstorms robots, re-using the code
generator used with the LBR robot was not feasible: the code generated
for the LBR produced too many artifacts for the Mindstorms robot's memory
to hold.  Instead, we developed a new code generator for the same RTS.
Due to the modularity of \lr, (a) integrating code generators is
straightforward and (b) the transformation from \lr models into SCs is
independent of subsequent code generation. Therefore, the new code
generator only translates SCs to Java. However, the current Mindstorms
version EV3 provides enough memory to re-use the same code generator as
used with the LBR.

\begin{ownfigure}
    \centering
    \includegraphics[width=.48\textwidth]{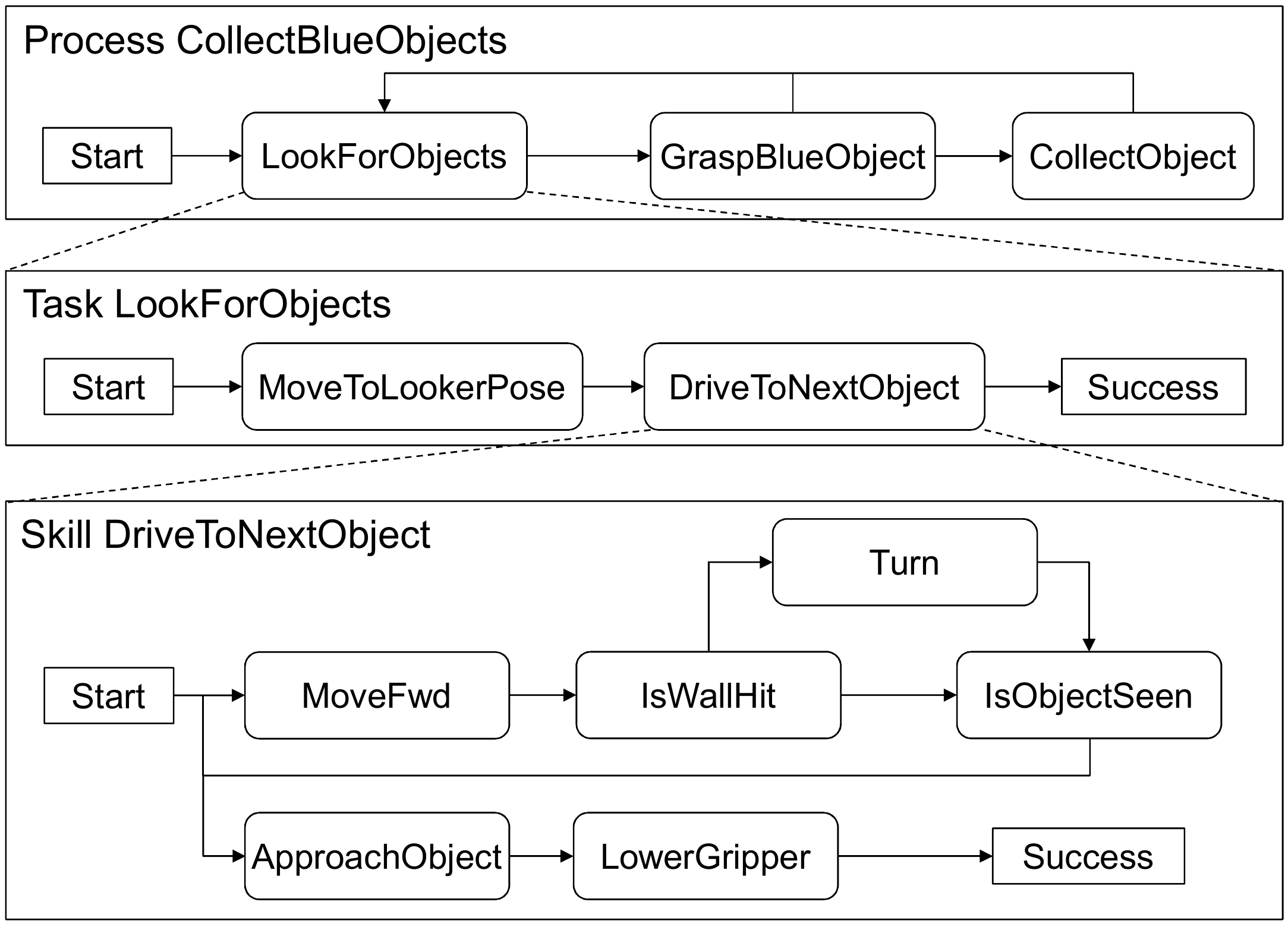}
    \caption{Process \code{CollectBlueObjects} with contained  
    task \code{LookForObjects} and skill \code{DriveToNextObject}.}
    \label{fig:ProcessCollectBlueObjects}
\end{ownfigure}

\subsection{KUKA iiwa Assembly Tasks}

We also applied \lr to assembly tasks with a KUKA iiwa robot in a case 
study with $10$ participants. The $3$ female and $7$ male participants 
were between $20$ and $59$ years old and had different degrees of 
expertise with model-driven engineering, robot programming, \lr, tablet 
computer usage and the iiwa robot. For instance, $60$\% of the 
participants had ``no'' previous experience with the iiwa robot and 
$20$\% had ``little'' previous experience with it. 

The participants were given a task and had to answer a questionnaire
afterwards. To fulfill the task, the participants had to pick up a
lightbulb from its initial position, move it to a thread, screw it into
the thread, and activate it via a switch. \autoref{fig:ScrewingTheLamp}
shows the initial setup with both lightbulb positions. To achieve this,
the participants were introduced to the concepts of \lr, the iiwa, and
tablet UI before they started modeling the task's solutions. The
introduction took between $10$ minutes and $1$ hour, depending on the
participant's previous knowledge.

\begin{ownfigure}
    \centering
    \includegraphics[width=.48\textwidth]{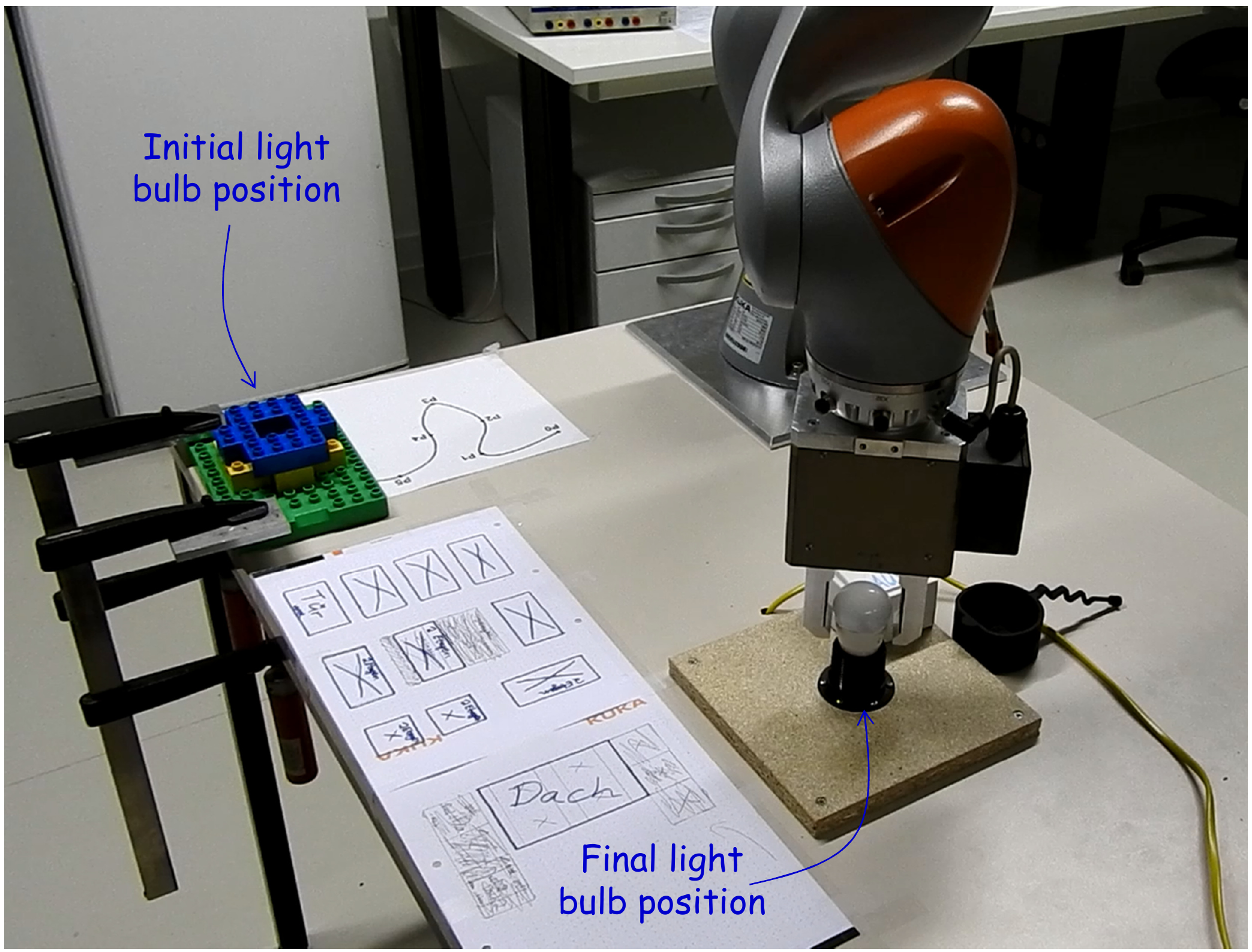}
    \caption{The case study setup with the lightbulb's initial
    position on top-left and the target thread bottom-right.}
    \label{fig:ScrewingTheLamp}
\end{ownfigure}

To model the required process and tasks, the participants used a
graphical editor displayed in \autoref{fig:EditorAtRuntime} on a tablet
computer. In this setup, all skills and actions were provided to the
participants, thus no robot API knowledge was required. This corresponds
with the idea, that the robot expert provides skills and tasks, and the
factory floor worker combines these only. In the end, all participants
completed the task. The fastest participant required $45$ minutes to
complete the complete case study, which included comprehending task 
description and available skills, teaching relevant poses to the robot, 
and solving the task. The slowest participant required $2$ hours.

\begin{ownfigure}
    \centering
    \includegraphics[width=.48\textwidth]{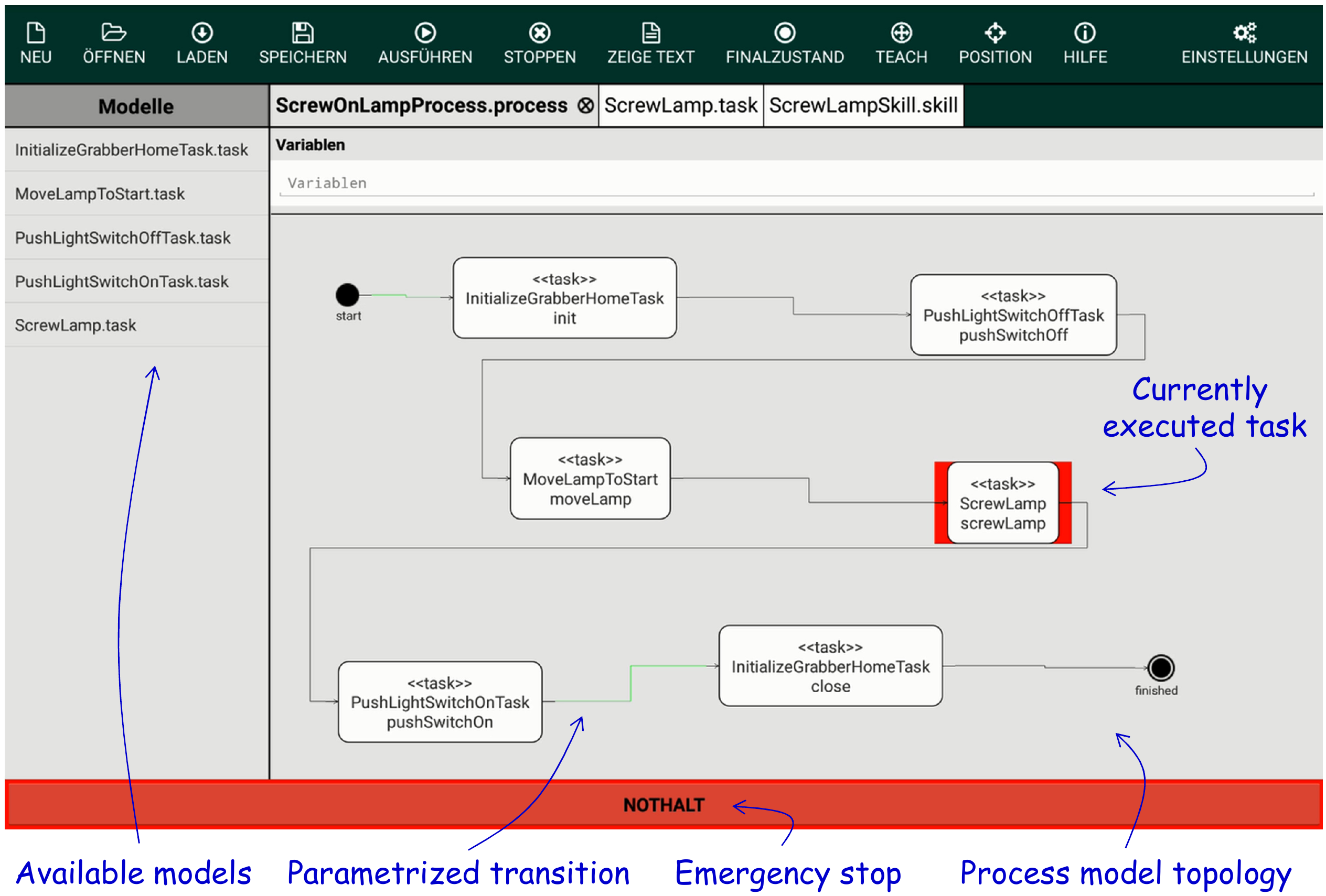}
    \caption{The tablet-based editor used to model the process and
    tasks to pickup, deliver, and screw the lightbulb.}
    \label{fig:EditorAtRuntime}
\end{ownfigure}

This case study was focused on applying \lr instead of developing
low level skills or actions for it and reflected its usage as intended
with factory floor workers. The study however is biased as $40$\% of the  
participants had at least mediocre or good programming skills. As
expected, these participants finished faster than the others. In
consequence, a future case study will work with participants with
little or no programming knowledge only. Nonetheless, even participants
without programming knowledge were confident as their feedback included
that \lr allowed ``easy robot programming'' and even ``enabled
untrained users to use robots as tools''.

\section{Discussion and Conclusion}
\label{sec:Conclusion}

\lr is a high-level robot programming toolchain feasible for both domain
experts and robot experts. Due to the abstraction of \lr,
processes and tasks can easily be re-used with different robots. Skills
and actions are tied to specific platforms but can easily be re-used for
different assembly processes. If the new platform represents the same kind of
robot (e.g. a LBR with seven degrees of freedom), only the parametrization of
the used actions needs to be reconfigured at model level. The \lr toolchain
furthermore enables re-use of code generators and run-time systems with compatible 
robots and supports development and execution of assembly tasks with powerful editors. 
Code generators can be re-used as long as target specific technical
restrictions, like available memory or target language are
fulfilled. The adaption of provided robot API has to be performed per robot /
API version used. Referring to \autoref{fig:Toolchain} the skill library
provider and the code generator developer need to perform target-specific
adaption, while the run-time system developer and the application modeler can
focus on a target-independent development. 

Case studies indicate that \lr helps to improve development of robotic
(assembly) processes. Nonetheless, the case studies pointed out issues:
currently, neither synchronous execution of tasks and skills, nor
parametrization of tasks with skills are supported. We will address the
issue of synchronous execution of tasks and skills and examine whether
such parametrization is useful without raising additional complexities.
Actions currently reference a single method of the robot's interface.
While facilitating re-use, this also leads to an increased number of
actions. In the future, we also will perform further case studies with
different robots and differently skilled users.

While the combined graphical and textual editor is helpful at modeling
tasks, it currently uses SWT to render tasks. Unfortunately, we have
experienced performance issues for large (more than 40 nodes) assembly
processes. We therefore will switch to a new rendering engine and we
will also examine whether on-line modeling as mentioned in
Sect.~\ref{sec:LR2} can be realized. Reasoning with tasks and skills
will be examined as well. With the strong formalism of preconditions and
postconditions, reasoning about the next action to be executed seems
useful to assist factory floor workers in modeling robot tasks.
Therefore we will examine how to integrate \lr with a model of the
environment, which, together with the actions, serves as the knowledge
base for assembly process reasoning.
With these models, the robot may react more dynamically to events and
thus increase the flexibility of assembly processes.


\bibliographystyle{IEEEtran}
\bibliography{paper}

\end{document}